\documentclass[aps,prl,reprint,groupedaddress]{revtex4-1}

\usepackage{graphicx}
\usepackage{amssymb,amsmath}
\usepackage{epstopdf}
\usepackage{float}
\usepackage{xcolor}
\usepackage{pifont}
\usepackage{bm}
\usepackage{mathrsfs}
\usepackage{bbold}

\newcommand{\br}{{\bf r}}

\newcommand{\beqa}{\begin{eqnarray}}
\newcommand{\eeqa}{\end{eqnarray}}

\renewcommand{\Im}{{\rm Im}}

\begin{document}

\title{Unification of quantum Zeno-anti Zeno effects and \\parity-time symmetry breaking transitions}

\author{Jiaming Li}
\email[]{lijiam29@mail.sysu.edu.cn}

\author{Tishuo Wang}

\author{Le Luo}
\email[]{luole5@mail.sysu.edu.cn} \affiliation{School of Physics and
Astronomy, Sun Yat-Sen University, Zhuhai, Guangdong, China 519082}

\author{Sreya Vemuri}

\author{Yogesh N Joglekar}
\affiliation{Department of Physics, Indiana University Purdue
University Indianapolis (IUPUI), Indianapolis, Indiana 46202, USA}

\date{\today}

\begin{abstract}
The decay of any unstable quantum state can be inhibited or enhanced
by carefully tailored measurements, known as the quantum Zeno effect
(QZE) or anti-Zeno effect (QAZE). To date, studies of QZE (QAZE)
transitions have since expanded to various system-environment
coupling, in which the time evolution can be suppressed (enhanced)
not only by projective measurement but also through dissipation
processes. However, a general criterion, which could extend to
arbitrary dissipation strength and periodicity, is still lacking. In
this letter, we show a general framework to unify QZE-QAZE effects
and parity-time ($\mathcal{PT}$) symmetry breaking transitions, in
which the dissipative Hamiltonian associated to the measurement
effect is mapped onto a $\mathcal{PT}$-symmetric non-Hermitan
Hamiltonian, thus applying the $\mathcal{PT}$ symmetry transitions
to distinguish QZE (QAZE) and their crossover behavior. As a
concrete example, we show that, in a two-level system periodically
coupled to a dissipative environment, QZE starts at an exceptional
point (EP), which separates the $\mathcal{PT}$-symmetric (PTS) phase
and $\mathcal{PT}$-symmetry broken (PTB) phase, and ends at the
resonance point (RP) of the maximum $\mathcal{PT}$-symmetry
breaking; while QAZE extends the rest of PTB phase and remains the
whole PTS phase. Such findings reveal a hidden relation between
QZE-QAZE and PTS-PTB phases in non-Hermitian quantum dynamics.
\end{abstract}

\maketitle

Quantum Zeno (anti-Zeno) effect is an important feature of a quantum
system, initially interpreted as that the evolution of the system
can be suppressed (enhanced) by measuring it frequently enough in
its known initial
state~\cite{Misra77,Schieve89,Itano90,Home97,Kaulakys97,Facchi00,
Lewenstein00,Streed06,Facchi01,Facchi10}. As an outgrowth of study
of the QZE (QAZE), they have been extensively used to control and
manipulate quantum systems, including changing the decay rate of an
unstable state~\cite{Fischer01,Yan13,Itano90,Zhang18}, protecting
quantum information~\cite{Gourgy18}, suppressing decoherence
~\cite{Streed06,Raimond12}, extending the lifetime of ultracold
molecular~\cite{Zhu14}, and suppressing the tunneling in an optical
lattice~\cite{Fischer01,Han09,Patil15}.

The meaning of QZE (QAZE) terms have since expanded, in which the
time evolution can be suppressed (enhanced) not only by projective
measurement but also by a variety of dissipation processes. Along
this line, a non-Hermitian term of dissipation $-i\gamma$ is
inserted into the Hermitian dynamics of the unstable system for
pulling away the curtain that hides the measurement
process~\cite{Schulman98,Muga08}. Such non-Hermitian Hamiltonians
are equivalent to the multi-mode Jaynes-Cummings model that includes
the coupling between the system and the measurement ``apparatus'',
enabling a unified theory to describe QZE (QAZE) through both
repetitive and continuous observations.


It is still an open question how to describe both measurement effect
and dissipation process so that deceleration (QZE) and
acceleration(QAZE) of the evolution of a quantum system can be
justified by a simple criterion. In Ref.~\cite{Kofman00}, such
criterion had been established for the case of frequent projective
measurement, in which the modified decay rate is simply determined
by the overlap of the reservoir coupling spectrum and the broadening
spectrum of the level that is frequently measured. However, a general
criterion, which could extend to dissipation with arbitrary small
strength, is still lacking. Such extension is not trivial, as
explained in Ref.~\cite{Kofman00}, the effect of projective
measurement is treated as phase randomization and induce level
broadening consequently. For small dissipation, we then need to deal
with partial decoherence which could induce the nontrival broadening of the quantum
level described by a stochastic, non-linear Liouville equation~\cite{Grigorescu98}.
Moreover, it is quite interesting to find out the
dependence of the strength of QZE (QAZE) on the realistic
parameters, especially at the limit of small dissipation strength
and arbitrary sequences, where the modified time evolution usually
do not behave as expected for idealized projective measurements.

Recently, emerging studies in passive $\mathcal{PT}$-symmetric
quantum systems indicate the appearance of a slow decay mode
associated with the PTB
phase~\cite{Li19,Joglekar18,Montiel18,Naghiloo19}. Based on these
discoveries, in this letter, we unify QZE (QAZE) and $\mathcal{PT}$
symmetry breaking transitions in non-Hermitian quantum mechanics,
and find $\mathcal{PT}$ symmetry breaking transitions play a general
role in determining QZE (QAZE) in open quantum systems. This
treatment enables us to search for QZE (QAZE) behaviors by analyzing
the phase diagram of a $\mathcal{PT}$ symmetric non-Hermitian
Hamiltonian. We explicitly show that $\mathcal{PT}$ symmetry
transitions hidden in a pure lossy two-level system could be used
for characterizing QZE (QAZE) precisely with arbitrary dissipation
strength and period.



The relation between QZE (QAZE) and $\mathcal{PT}$-symmetry
transition is built as follows: First, the projective measurement
related to QZE (QAZE) is considered as a pure loss term that couples
the system to the environment at the strong limit of dissipation
strength. Then, by decreasing the dissipation strength, QZE (QAZE)
is studied in the weak dissipation regime. This non-Hermitian
Hamiltonian is given by
\begin{equation}
\label{eq:zeno1}
H=\omega_0\lvert 1\rangle\langle 1\rvert+ (\omega -
2i \gamma)a^{\dagger}a +a^{\dagger}\Phi\lvert 0\rangle\langle
1\rvert+a\Phi^{*}\lvert 1\rangle\langle 0\rvert\,,
\end{equation}
which describe a system that an atom in the excited state $\lvert
1\rangle$ of energy $\omega_0$ decays to the ground state of zero
energy by emitting photons, falling in a single photon mode with
frequency $\omega$. We model the action of the measurement by
considering that the state to which $\lvert 1\rangle$ decays is
itself unstable with a decay rate $\gamma$~\cite{Schulman98}.
Second, the pure dissipative Hamiltonian associated to QZE (QAZE)
can be mapped into a balanced gain-loss Hamiltonian with
$\mathcal{PT}$ symmetry. Defining $\omega_a=(\omega_0+\omega)/2$ and
$\Delta=(\omega_0-\omega)/2$, we have $H=H_0+H_{int}$, where
$H_0=(-i\gamma+\omega_a)\mathbb{1}+\Delta\sigma_z$ with $\mathbb{1}$
is the unit matrix and $\sigma_z$ the Pauli matrix, and we have
\begin{equation}
\label{eq:Zeno2}
H_{int}=\begin{pmatrix}
+i\gamma &\Phi^{*} \\
\Phi & -i\gamma
\end{pmatrix}
\end{equation}
$H_{int}=\mathcal{PT}H_{int}\mathcal{PT}$ is a balanced gain-loss
Hamiltonian $H_{PT}$ remaining invariant under $\mathcal{PT}$
operation with the parity operator given by $\mathcal{P}=\sigma_x$
and the antiliner time-reversal operator given by
$\mathcal{T}i\mathcal{T}=-i$. This $\mathcal{PT}$-symmetric
Hamiltonian allows phase transitions in its eigenvalue spectrum,
where the eigenvalue changes from purely real to complex-conjugate
pairs. Known as passive $\mathcal{PT}$-symmetry transitions, the
transitions of a no-gain system with mode-selective losses has been
successfully observed in various systems~\cite{Guo09,
Ruter10,Regensurger12,Heiss12,Schindler11,Assawaworrarit17,Chitsazi17,Montiel18,
Peng13,Xu16,Brandstetter14,XuJian17,Li19,Xiao17,Wu19,Naghiloo19}.
Third, we adopt the frequency-dependent method to signal the QZE
(QAZE)~\cite{Zhang15,Chaudhry16, Chaudhry14,Wu17}. In the
frequency-dependent picture, QZE(QAZE) is defined in such a way that
the effective decay rate $\Gamma(\omega)$ decreases (increases) as
the measurement (dissipation) frequency $\omega$ increases. This
definition provides a clear physical picture: the rapidly repeated
measurements suppressed (enhanced) the relaxation process of the
unstable state, thus leading to the QZE (QAZE).

Based on these ideas, we put our focus on analyzing how the strength
and periodicity of the dissipation term $\gamma$ play a role in the
decay rate $\Gamma$ of an unstable system, especially on the slow
decay mode that could induce QZE (QAZE) when tuning the strength or
frequency of the dissipation. For a static dissipation, at small
dissipation strength, the decay rate of the system
$\Gamma(\gamma_0)$ increases as $\gamma_0$, whereas the decay rate
could be slowed down by increasing $\gamma_0$ at larger strength.
For a time-periodically modulated dissipation $\Gamma(\omega)$, the
decay rate depends on both the magnitude of the dissipation and its
modulation frequency, leading to a rich phase diagram separated by
multiple PTS and PTSB phases~\cite{Joglekar14,Li19}. In the PTS
phase, the decay rate always increases as the increase of the
dissipation frequency. In contrast, the eigenvalues have two
branches in the PTSB phase, one is the ``slow mode'' with less
imaginary(loss) components than the other ``fast mode''. While the
fast mode decays quickly, the slow mode survives in the longer time
and dominate the time evolution so that the effective decay rate is
slowed down.

\begin{figure}[h]
\centering
\includegraphics[width=\columnwidth]{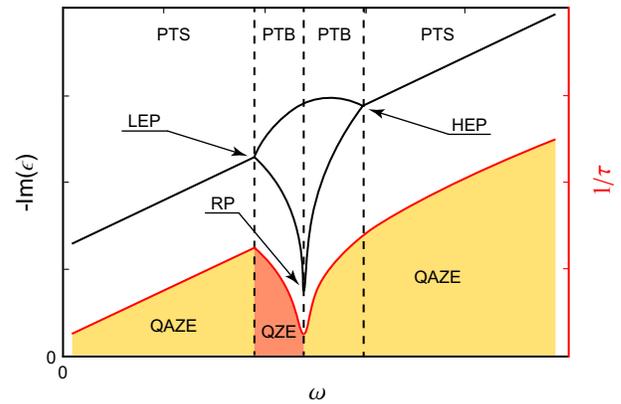}
\caption{The concept picture indicating the relation between QZE
(QAZE) and $\mathcal{PT}$-symmetry breaking transitions. Black solid
line: the dependence of the imaginary part of the quaienergy
$-\Im(\epsilon)$ of a passive $\mathcal{PT}$-symmetry Hamiltonian on
the frequency of the dissipation $\omega$. Red solid line: The
effective decay rate of an unstable system with the characteristic
life time $\tau$. QZE (QAZE) represents for quantum zeno (anti-zeno)
effect. PTS (PTB) represents for $\mathcal{PT}$-symmetric
($\mathcal{PT}$-symmetry broken) phase. LEP(HEP) is the exceptional
point of $\mathcal{PT}$-symmetry breaking transitions with PTS at
low (high) frequency side. RP is the resonant point of the PTB
phase. \label{Fig1}}
\end{figure}

It is naturally to ask if a PTB phase corresponds to QZE while a PTS
phase to QAZE. The answer is that the direct correspondence is not
valid. Instead, we present a general relation shown in
Fig.~\ref{Fig1} by analyzing the dependence of the imaginary part of
the eigenvalues on the frequency of the dissipation. In the PTS
phase, the imaginary part of the eigenvalue increases as the
dissipation frequency $\omega$ becomes larger, so that QAZE will be
observed. When the dissipation frequency is larger than the LEP, the
``slow mode'' of the eigenvalue appears and the imaginary part of
``slow mode'' decreases as $\omega$ increases. In the PTB regime,
the decay rate of the system is dominated by the ``slow mode'' which
inhibits the decay, indicating QZE until $\omega$ increases to the
RP. Above that, the imaginary part of ``slow mode'' increases as
$\omega$ increases, showing QAZE. The QAZE exists in the PTB phase
that is above the resonant point, and remains in the PTS phase while
$\omega$ increases and passed the HEP. This analysis uses the
frequency response of a decay system to define QZE (QAZE) and
depends on the eigengmode behavior of a passive
$\mathcal{PT}$-symmetric system. These arguments are rather general
not depending on the details of Hamiltonian, so we think this is a
universal relation to unify parity-time symmetry breaking
transitions of a non-Hermitian Hamiltonian and QZE (QAZE) in open
quantum systems.



To make physics clear in a simple context, we illustrate the above
relation using a two-level dissipative Rabi system driven by the
resonant photon mode
$H_L=-J(t)(\lvert\uparrow\rangle\langle\downarrow\rvert+\lvert\downarrow\rangle\langle\uparrow\rvert)
-2i\gamma(t)\lvert\downarrow\rangle\langle\downarrow\rvert $, in
which the coupling rate $J(t)$ and the dissipation rate $\gamma(t)$
of the $\lvert\downarrow\rangle$-level are both time dependent. As
shown in Eq.~\ref{eq:Zeno2}, this Hamiltonian can be written as
$H_L=-i\gamma \mathbb{1}+H_{PT}$.
\begin{figure}[htbp]
\includegraphics[width=1\columnwidth]{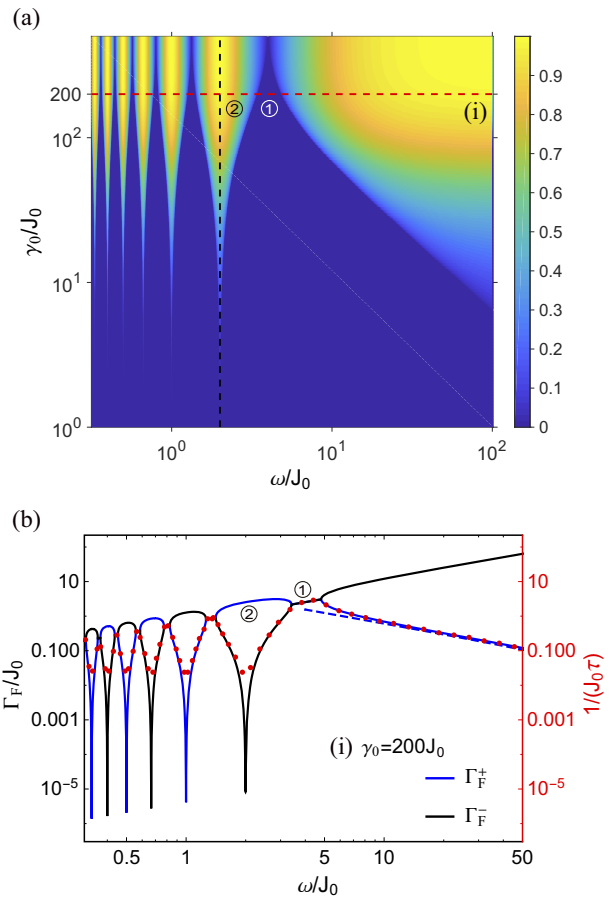}
\caption{The decay rates of a two-level system with a time-dependent
dissipation modeled by $\mathcal{PT}$-symmetric Hamiltonian. (a) The
phase diagram of $\mathcal{PT}$ symmetry breaking, in which the
color presents $\mu(\gamma_0,\omega)$. The vertical axis is the
normalized dissipation amplitude $\gamma_0/J_0$, and the horizontal
axis is normalized modulation frequency $\omega/J_0$. Note that the
phase diagram is obtained by fixing $\tau_1$ and varying $T$.
Although the widths of the PTS phases (deep blue color) and PTB
phases (all other colors) depend on $\tau_1$, the structure of the
phase diagram as well as the location of the resonance peaks (one of
them represented by the black-dash line) does not depend on
$\tau_1$. (b) The comparison of the decay rates $\Gamma_F$ and the
lifetime $\tau$ along the red-dashed line in Fig~\ref{Fig2}(a) with
$\gamma_0=200J_0$. The red dot is the numerical simulation of the
lifetime of the unstable state. \ding{172} (\ding{173}) presents one
of the PTS (PTB) regime. \label{Fig2}}
\end{figure}

To reveal QZE (QAZE), $J_0$ is constant, and a square-wave
modulation of the dissipation $\gamma(t)$ with pulse width $\tau_1$
and period $T$ is applied
\begin{equation}
\label{eq:Floquet}
\gamma(t)=\left\{\begin{array}{ccc}
\gamma_0 & & 0\leq t < \tau_1,\\
0 & & \tau_1 \leq t < T,\\
\end{array}\right.
\end{equation}
where $T=2\pi/\omega$ is the period of the Hamiltonian i.e.
$H_L(t+T)=H_L(t)$. The PTS (PTB) phases are defined via
quasienergies $\epsilon^{\pm}_F$ of the effective Floquet
Hamiltonian $H_L$, which are obtained from the eigenvalues of the
non-unitary time evolution operator for one period $G^\prime(T)$
(see Supplementary Materials A). Here
$G^\prime(T)=e^{-iH_L(T-\tau_1)}e^{-i H_L\tau_1}=e^{-\gamma_0
\tau_1}G(T)$ with $G(T)=e^{-iH_{PT}(T-\tau_1)}e^{-i H_{PT}\tau_1}$
as the time evolution operator of balanced gain and loss.
$\epsilon^{\pm}_F$ is then given by
\begin{equation}
\epsilon_F^{\pm}=-i\gamma_0\tau_1/T+i\ln(\Lambda_F^{\pm})/T
\label{eq:epsilonF}
\end{equation}
with $\Lambda_F^{\pm}$ of the eigenvalue of $G(T)$,
\begin{eqnarray}
\label{eq:Lambda}
\Lambda_F^{\pm}&=&c_{1}c_{2}-\frac{J_0}{\epsilon_0}s_{1}s_{2}\nonumber\\
&\pm&\sqrt{(\frac{\gamma_0}{\epsilon_0}s_{1})^{2}-(c_{1}s_{2}+\frac{J_0}{\epsilon_0}s_{1}c_{2})^{2}}.
\end{eqnarray}
The parameters are defined as
$c_{1}\equiv\cosh(\epsilon_0\tau_{1})$,
$c_{2}\equiv\cos[J_0(T-\tau_1)]$,
$s_{1}\equiv\sinh(\epsilon_0\tau_{1})$,
$s_{2}\equiv\sin[J_0(T-\tau_1)]$ and
$\epsilon_0\equiv\sqrt{\gamma^2_0-J^2_0}$.

The imaginary parts of the quasi-energies
$\epsilon^{\pm}_F(\gamma_0,\omega,\tau_1)$ determine the decay rates
$\Gamma^{\pm}_F=-2\,\Im\epsilon^{\pm}_F$. In the PTS phase,
$\Lambda_F^{\pm}$ are complex conjugates of each other and have the
same magnitude. Therefore, the real part of $\ln\Lambda_F^{\pm}$ are
the same, so does the imaginary part of $\epsilon_F^{\pm}$. Thus the
decay rates $\Gamma^{\pm}_F$ are equal and increase when $\gamma_0$
increases. In the PTB phase, both $\Lambda_F^{\pm}$ become purely
real, leading to the imaginary parts of $\epsilon_F^{\pm}$ different
and the emergence of two different decay rates, named as ``slow
mode'' and ``fast mode''. Two modes arises at the exceptional point
of the $\mathcal{PT}$ symmetry breaking transition. The degree of
symmetry breaking is described by a dimensionless parameter
$\mu(\gamma_0,\omega)=||e^{-i\epsilon_F^{+}T}|-|e^{-i\epsilon_F^{-}T}||$.
As an example, Figure~\ref{Fig2}(a) shows $\mu(\gamma_0,\omega)$ for
dissipation with the pulse parameter of $J_0\tau_1=0.01$, and
Figure~\ref{Fig2}(b) shows the decay rates $\Gamma_F$ obtained along
the red-dash line in Fig~\ref{Fig2}(a). The coincidence between the
lifetime of the unstable state and the decay rates of the eigenmodes
have been confirmed from large to small dissipation strength
(Supplementary Materials C).

We could extend this result to the whole $\mathcal{PT}$ phase
diagram as shown in Fig.~\ref{Fig2}. There are multiple PTS and PTB
blocks with the resonant frequencies of PTB as $\omega_n/J_0=
2/n$, where $n=1,2,3...$(see Supplementary Materials B). In one of
the PTS blocks (marked as \ding{172}) shown in Fig.~\ref{Fig2},
$\Gamma_{F}^{\pm}$ decreases with the decreases of the modulation
frequency, indicates QAZE. As the modulation frequency decreases,
the system experiences a phase transitions from the PTS to PTB.
After crossing the exceptional point, in one of the PTB blocks
(marked as \ding{173}), the decay rate of the slow mode is not
monotonous. Below the PTB resonance, the $\Gamma_{F}^{-}$ decreases
with the increase of $\omega$ and reverses trend above the
resonance, so both the QZE and QAZE appear in the PTB phase, and the
transition from the QAZE to the QZE is determined by
$\omega_{n}$.

This framework leads to the unification of the $\mathcal{PT}$
symmetry braking transition and QAZ(QAZE). But does this unification
also support the results of the projective measurements and
continuous observation (static dissipation)? The answer is yes, and
we confirm that the universality of this unification is applied for
both cases.

First, projective measurements. For comparing with QZE(QAZE) induced
by projective measurements, we need to consider the limit of the
large and frequent dissipation with $\gamma_0/J_0\gg 1$ and
$\omega/J_0\gg 1$, the decay rate of the states $\Gamma_{F}^{\pm}$
is simplified as
$\Gamma_{F}^{\pm}=(J_0^{2}/\gamma_0)(\tau_{1}/T)+2\ln((c_2\pm\sqrt{1-s_2^2})/2)/T$.
With $\omega/J_0\gg 1$, $\cos(J_0\tau_{2})\rightarrow
e^{-(J_0\tau_{2})^2/2}$, we got
\begin{equation}
\label{eq:5}
\Gamma_{F}^{+}=\frac{J_0^{2}}{\gamma_0}\frac{\tau_{1}}{T}+J_0^{2}\frac{\tau_{2}^{2}}{T}
\end{equation}
On the other hand, a two-level system in which the projective
measurements applied to the final state allows QZE ~\cite{Streed06},
in which the two-level system is driven at Rabi frequency $\omega_R$
with the initial population of the atoms in the
$\lvert\uparrow\rangle$, while the final dissipative state
$\lvert\downarrow\rangle$ has a decay rate $\gamma_c$ coupled to the
third state. When $N$ rapid projective measurements with time
intervals of $\delta t=t/N$ are applied the final state, the
survival probability of the initial state after N times measurements
is $p^N(t)=p(\delta t)^N\simeq [1-(\omega_R\delta t/2)^2]^N$. When
$\delta t\ll \pi/\omega_R$, the decay rate of $p^N(t)$ is
$1/\tau_{QZE}= \omega_R^2 \delta t/4 $, showing that the projective
measurements slow down the decay of the state. In the real
experiments, the measurement time is finite, both the measurement
pulse duration $t_p$ and the time interval between the two
consecutive pulses $\delta t$ are needed to be considered, giving
the decay rate
\begin{equation}
\label{eq:6}
\frac{1}{\tau}=\frac{\omega_R^{2}}{\gamma_c}\frac{t_p}{t_p+\delta
t}+\frac{\omega_R^{2}}{4}\frac{\delta t^{2}}{t_p+\delta t}
\end{equation}
It is obvious that Eq.~\ref{eq:5} and Eq.~\ref{eq:6} are equivalent,
which indicates that QZE(QAZE) can be well understood in terms of
the picture of $\mathcal{PT}$ symmetry breaking transition. The
blue-dash line in Fig.~\ref{Fig2} (b) is plot as Eq.~\ref{eq:6},
showing that the two decay rates are in agreement with each other
very well.

\begin{figure}[h]
\centering
\includegraphics[width=\columnwidth]{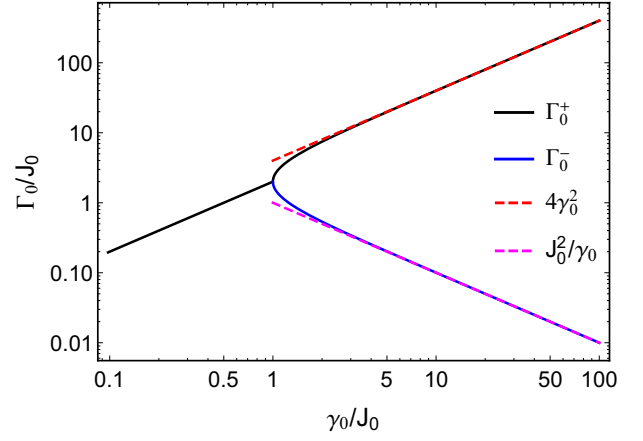}
\caption{The decay rates of continuous observation as a function of
the static dissipation. The black (blue) lines present the two
eigenmode $\Gamma_0^{+} (\Gamma_0^{-})$ in the PTB phase respectively
(the two eigenmodes overlap in the PTS phase). The dashed lines present the decay rates
at the limit of $\gamma_0 \gg J_0$.}
\label{Fig3}
\end{figure}

Second, continuous observation (static dissipation). In the static
case, the eigenvalues are given by
$\lambda_{\pm}=-i\gamma_0\pm\sqrt{J_0^2-\gamma_0^2}$ where
$J_0,\gamma_0$ are the static parameters. The decay rates
$\Gamma_0^\pm$ as a function of $\gamma_0/J_0$ is shown in
Fig.~\ref{Fig3}. When $\gamma_0\leq J_0$ with the $J_0$ as the LEP,
the system is in the PTS phase and the decay rates of the two
eigenmodes are equal, $\Gamma_0^{\pm}=2\gamma_0$. In this phase, the
decay rate increases as $\gamma_0$ increases. When $\gamma_0>J_0$,
the system is in the PTB phase, leading to the emergence of two
modes given by
\begin{eqnarray}
\label{eq:lifetime_static}
\Gamma_0^{-}&=&2(\gamma_0-\sqrt{\gamma_0^2-J_0^2}) \underset{\gamma_0\gg J_0}{\rightarrow} \frac{J_0^2}{\gamma_0},\nonumber\\
\Gamma_0^{+}&=&2(\gamma_0+\sqrt{\gamma_0^2-J_0^2})\underset{\gamma_0\gg
J_0}{\rightarrow}4\gamma_0.
\end{eqnarray}
In the limit $\gamma_0\gg J_0$, the decay rate for the ``fast mode''
doubles, $\Gamma_0^+\rightarrow 4\gamma_0$, whereas that for the
``slow mode'' vanishes, $\Gamma_0^-\rightarrow J_0^2/\gamma_0$.
These values coincide the continuous QZE case with theory in
Ref.~\cite{Schulman98} and experiment in Ref.~\cite{Streed06}. The
populations of up and down levels decays are given by
$1/\tau_{\lvert\uparrow\rangle}=\omega_R^2/\gamma_c\nonumber$ and
$1/\tau_{\lvert\downarrow\rangle}=\gamma_c$ from the picture of
quantum measurement. It is clear that the two approaches are
equivalent provided $\omega_R=2J_0$ and $\gamma_c=4\gamma_0$. In the
strong-dissipation limit $\gamma_0\gg J_0$, the slowly-decaying
eigenmode has a near-unity overlap with the $\lvert\uparrow\rangle$,
while the rapidly-decaying eigenmode is mostly aligned with
$\lvert\downarrow\rangle$. Thus, the PTB phase provides a suitable
generalization of the continuous QZE when the dissipation strength
is moderate, $\gamma_0\gtrsim J_0$. On the other hand, when
$\gamma_0\leq J_0$, the two decay rates $\Gamma_0^\pm=2\gamma_0$
increase with increasing dissipation, which is consistent with the
QAZE.

We formulate a general picture of QZE(QAZE) in the two-level
dissipative Rabi system based on the phase diagram of $\mathcal{PT}$
symmetry. QZE is always observed, above certain modulation frequency
$\omega$, in the strong dissipative regime $\gamma_0/J_0\gg 1$. But,
even deep in the strong dissipation regime, QAZE could also be
observed around specific modulation frequencies near the EP points.
On the other hand, at small dissipation strengths, QAZE can be
observed at most modulation frequencies. Conversely, the QZE regime
survives down to vanishingly small dissipation strengths, i.e.
$\gamma_0/J_0\ll 1$, only with the modulation frequencies in the
range between the LEP and the RP of the PTB phase.


In conclusion, we unify the symmetry transitions associated to
$\mathcal{PT}$-symmetric non-Hermitian Hamiltonians with quantum
measurement effect of QZE (QAZE). Using a dissipation term, instead
of projective measurement, the interaction between the unstable
system and the environment can be either strong or weak, enabling a
systematic study of weak dissipation, in which QZE(QAZE) will not be
as manifest as in the projective measurement case. We find that
$\mathcal{PT}$ phase transitions exit in all types of QZE(QAZE)
effects whether the dissipation is strong or weak, periodically or
static. Such findings helps to explore QZE(QAZE) physics in more
complex setups, such as beyond Markovian
approximation~\cite{Weiwu17,Zhou17} and with many-body
interactions~\cite{Yaodong18,Heinrich19}, with a simple,
quantitative criterion, leading further studies of the deep
relations between quantum measurement effects and the dynamics of
non-Hermitian open systems.


\begin{acknowledgments}
JLi received supports from National Natural Science Foundation of
China (NSFC) under Grant No.11804406, Fundamental Research Funds for
Sun Yat-sen University, Science and Technology Program of Guangzhou
2019-030105-3001-0035. YNJ was supported by NSF grant DMR-1054020.
LL received supports from NSFC under Grant No.11774436, Guangdong
Province Youth Talent Program under Grant No.2017GC010656, Sun
Yat-sen University Core Technology Development Fund, and the
Key-Area Research and Development Program of GuangDong Province
under Grant No.2019B030330001.
\end{acknowledgments}


\bibliography{tail1908pt}
\end{document}